\documentclass[conference]{IEEEtran}
\makeatletter
\def\endthebibliography{%
  \def\@noitemerr{\@latex@warning{Empty `thebibliography' environment}}%
  \endlist
}
\makeatother
\usepackage{trfsigns}
\usepackage{subfig}
\usepackage{graphicx}
\usepackage{float}
\usepackage{tikz}
\usepackage[nolist]{acronym} 
\usepackage{pgfplots}
\usepackage{circuitikz}
\pgfplotsset{compat=newest}
\usepackage{units}
\usepackage[group-separator={,}]{siunitx}
\usepackage{amsmath, amsbsy, amssymb}
\definecolor{darkgreen}{rgb}{0.125,0.5,0.169}
\usepackage{xcolor}
\definecolor{mittelblau}{RGB}{0, 126, 198}
\definecolor{violettblau}{cmyk}{0.9, 0.6, 0, 0}
\definecolor{rot}{RGB}{238, 28 35}
\definecolor{apfelgruen}{RGB}{140, 198, 62}
\definecolor{gelb}{RGB}{255, 229, 0}
\definecolor{orange}{RGB}{244, 111, 33}
\definecolor{pink}{RGB}{237, 0, 140}
\definecolor{lila}{RGB}{128, 10, 145}
\definecolor{hellgrau}{RGB}{224, 224, 224}
\definecolor{mittelgrau}{RGB}{128, 128, 128}
\definecolor{dunkelgrau}{RGB}{80,80,80}
\definecolor{anthrazit}{RGB}{19, 31, 31}

\IEEEoverridecommandlockouts

\begin{document}

%\section*{To do}
%\begin{itemize}
%\item Attention: SPAWC has only 4 pages + 1 page of references %(see: $http://spawc2018.org/files/Call_for_papers_1009.pdf$)
%\end{itemize}

%%%%%%%%%%Paper starts here%%%%%%%%%%%

%\title{A novel Massive MIMO Channel Sounder for a comprehensive CSI Database}
%\title{Deep Learning-based Indoor Positioning using Massive MIMO Channel Sounding Data}
\title{Novel Massive MIMO Channel Sounding Data applied to Deep Learning-based Indoor Positioning}

\author{\IEEEauthorblockN{ Maximilian Arnold\IEEEauthorrefmark{1}, Jakob Hoydis\IEEEauthorrefmark{2}, and Stephan ten Brink\IEEEauthorrefmark{1}}
\IEEEauthorblockA{\IEEEauthorrefmark{1}Institute of Telecommunications, University of  Stuttgart, 70659 Stuttgart, Germany 
    \\\{arnold,tenbrink\}@inue.uni-stuttgart.de
}
\IEEEauthorblockA{\IEEEauthorrefmark{2}Nokia Bell Labs, Route de Villejust, 91620 Nozay, France
\\jakob.hoydis@nokia-bell-labs.com
}
}

 \maketitle

\begin{acronym}
 \acro{CSI}{channel state information}
 \acro{TDD}{Time Division Duplexing}
 \acro{FDD}{Frequency Division Duplexing}
 \acro{ECC}{error-correcting code}
 \acro{MLD}{maximum likelihood decoding}
 \acro{HDD}{hard decision decoding}
 \acro{IF}{intermediate frequency}
 \acro{RF}{radio frequency}
 \acro{SDD}{soft decision decoding}
 \acro{NND}{neural network decoding}
 \acro{ML}{maximum likelihood}
 \acro{GPU}{graphical processing unit}
 \acro{BP}{belief propagation}
 \acro{LTE}{Long Term Evolution}
 \acro{BER}{bit error rate}
 \acro{SNR}{signal-to-noise-ratio}
 \acro{ReLU}{rectified linear unit}
 \acro{BPSK}{binary phase shift keying}
 \acro{AWGN}{additive white Gaussian noise}
 \acro{MSE}{mean squared error}
 \acro{LLR}{log-likelihood ratio}
 \acro{MAP}{maximum a posteriori}
 \acro{NVE}{normalized validation error}
 \acro{BCE}{binary cross-entropy}
 \acro{BLER}{block error rate}
 \acro{SQR}{signal-to-quantisation-noise-ratio}
 \acro{MIMO}{multiple-input multiple-output}
 \acro{OFDM}{orthogonal frequency division multiplex}
 \acro{RF}{radio frequency}
 \acro{LoS}{line of sight}
 \acro{NLoS}{non-line of sight}
 \acro{NMSE}{normalized mean squared error}
 \acro{CFO}{carrier frequency offset}
 \acro{SFO}{sampling frequency offset}
 \acro{IPS}{indoor positioning system}
 \acro{TRIPS}{time-reversal IPS}
 \acro{RSSI}{received signal strength indicator}
 \acro{MIMO}{multiple input-multiple output}
 \acro{ENoB}{effective number of bits}
 \acro{AGC}{automated gain control}
 \acro{ADC}{analog to digital converter}
 \acro{ADCs}{analog to digital converters}
 \acro{FB}{front bandpass}
 \acro{FPGA}{field programmable gate array}
 \acro{JSDM}{Joint Spatial Division and Multiplexing}
 \acro{NN}{Neural Network}
 \acro{IF}{intermediate frequency}
 \acro{DSP}{digital signal processing}
 \acro{AFE}{analog front end}
 \acro{SQNR}{signal-to-quantisation-noise-ratio}
 \acro{ENoB}{effective number of bits}
 \acro{AGC}{automated gain control}
 \acro{PCB}{printed circuit board}
 \acro{EVM}{error vector magnitude}
 \acro{CDF}{cumulative distribution function}
 \acro{MRC}{maximum ratio combining}
 \acro{DL}{deep-learning}
\end{acronym}

%\tikzexternaldisable

\begin{abstract}
With a significant increase in area throughput, Massive MIMO has become an enabling technology for fifth generation (5G) wireless mobile communication systems. Although prototypes were built, an openly available dataset for channel impulse responses to 
verify assumptions, e.g. regarding channel sparsity, is not yet available. In this paper, we introduce a novel channel sounder architecture, capable of measuring multi-antenna and multi-subcarrier channel state information (CSI) at different frequency bands, antenna geometries and propagation environments. 
The channel sounder has been verified by evaluation of channel data from first measurements. 
Such datasets can be used to study various \ac{DL} techniques in different applications, e.g., for indoor user positioning in three dimensions, as is done in this paper. 
Not only we do achieve an accuracy better than \SI{75}{\centi \metre} for \ac{LoS}, as is comparable to state-of-the-art conventional positioning techniques, but also obtain similar precision for the more challenging case of \ac{NLoS}. 
Further extensive indoor/outdoor measurement campaigns will provide a more comprehensive
open CSI dataset, tagged with positions, for the scientific community to further test various algorithms.
\end{abstract}

\acresetall

\section{Introduction}
An over-provisioning of antennas in Massive MIMO systems can be used to create orthogonal channels in space, resulting in higher area throughput \cite{Bjrnson2017}. Therefore, Massive MIMO has become an attractive preoposition for fifth generation (5G) wireless mobile communication systems \cite{5GTechCommMag,MMIMO5GCommMag}.
Although Massive MIMO prototypes were available as early as 2012 \cite{Xie2015HekatonEA,Shepard2012ArgosPM,7931558} and were showing promising results regarding the expected area throughput, an open source \ac{CSI} database, tagged with GPS coordinates, including different frequency bands, antenna geometries and propagation environments to better evaluate the performance of Massive MIMO, is not yet available. Such publicly available datasets can be used for, e.g., testing precoding schemes with actual channel data, and to verify predictions on the achievable sum-rate. Thus, a novel channel sounder which is not limited to specific frequency bands or antenna geometries is needed for creating such database.

It is widely acknowledged in the research literature that Massive MIMO favors \ac{TDD} as mode of operation, owing to the fact that the piloting overhead is independent of the number of basestation antennas.
This mode is also referred to as ``canonical'' Massive MIMO  \cite{Bjrnson2017}.
Still, industry strives for enabling \ac{FDD} Massive MIMO, as current network architectures favor \ac{FDD} systems \cite{7402270}. Thus, there is a significant body of papers assuming a (still to be verified) sparse channel model, leading to the possibility of using \ac{JSDM} \cite{6310934} and compressive sensing techniques \cite{8284057}. By exploiting channel sparsity, different algorithms show the potential of achieving the same spectral efficiency as canonical (i.e., TDD-based) Massive MIMO  \cite{7008286}. However the question remains whether the measured channels in the lower frequency bands from 1 to \SI{6}{\giga \hertz} exhibit the property of sparsity in  real-world propagation environments. 
%Following different requirements for channel properties and to verify those on actually measured channels, the necessity for measuring different \ac{CSI} with the corresponding scenario and positions is increased. 
Due to the many antennas in Massive MIMO systems, measurement campaigns create a large amount of data, suggesting that machine learning and, in particular, \ac{DL} techniques can be improved by making use of those datasets, e.g., for studying the unsolved task of indoor user localization
\cite{Arnold2018OnDL,Lymberopoulos2018,Decurninge2018CSIbasedOL,Vieira2017DeepCN}: A \ac{NN} could exploit the large number of different channel estimates (per antenna and subcarrier) to predict the current position of the user.
Just from considering these two possible applications, the need for a publicly available \ac{CSI} position-tagged database becomes obvious.

In this paper, the main requirements for a channel sounder are shown and a novel low cost architecture for Massive MIMO measurements is proposed. Compare to the channel sounder proposed in \cite{ChannelSounderforMassiveMIMO} or any yet to be seen architecture, the proposed channel sounder reduces the number of \ac{ADC} per RF chain and therefore RF-topology complexity. The flexibility of the channel sounder regarding different frequency bands, antenna geometries and propagation scenarios is discussed.
Also, sanity checks regarding viability, stability, hardware impairments and reproducibility of the channel sounding results are presented. 
As an application of the sounding data, initial results based on \ac{DL} for indoor user positioning are shown, where an accuracy of better than \SI{75}{\centi \metre} in three dimensions is achieved, indicating the potential of using Massive MIMO for indoor user positioning exploiting actual channel measurements.

\section{Channel Sounder Architecture}
Let us first introduce the basic functionalities and requirements for the channel sounder is presented which is later used for verification of the proposed architecture. 

\begin{figure}[]
	\centering
	\includegraphics[width=0.5\textwidth]{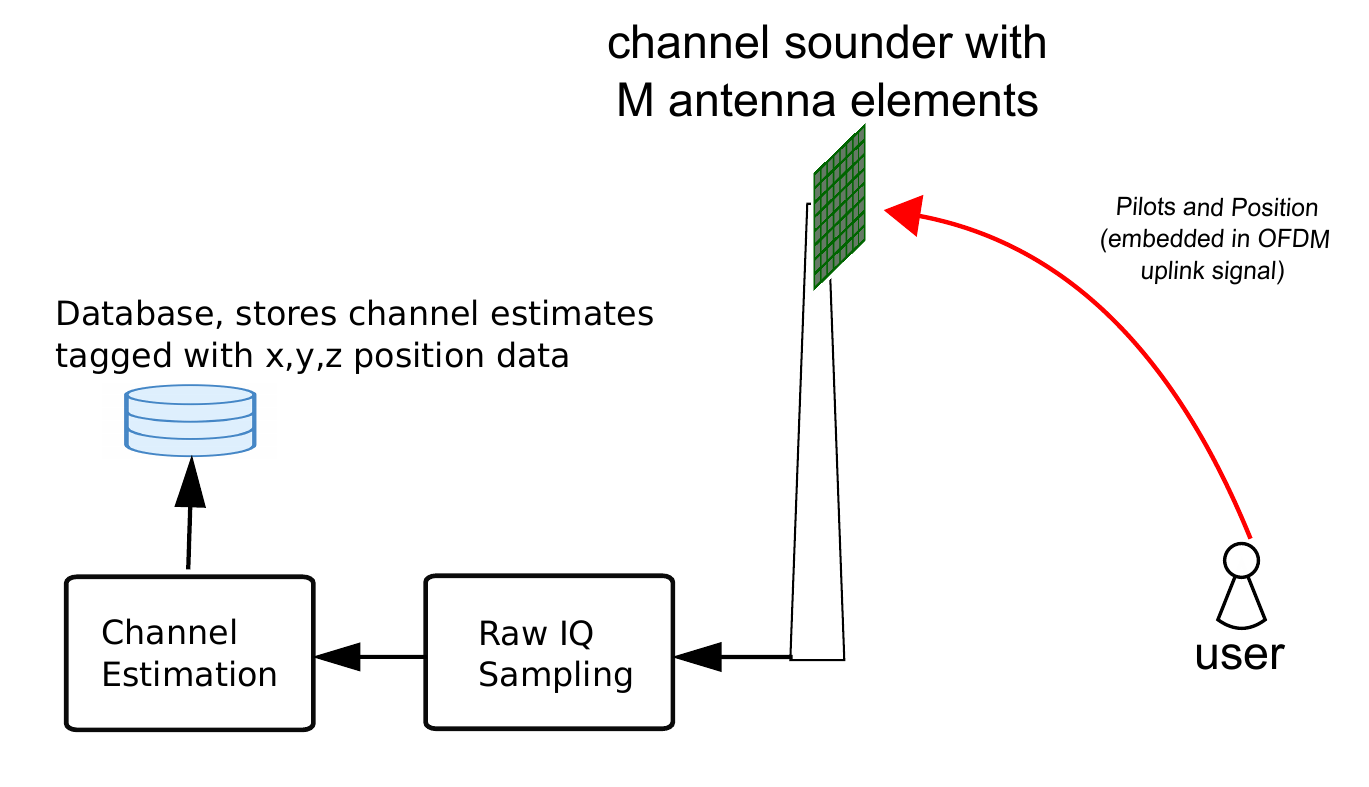}
	\caption{Massive MIMO channel sounder (CSI, user position)}
	\label{fig:Principle_Channel_Sounder}
        
\end{figure}

Fig. \ref{fig:Principle_Channel_Sounder} shows the system model of a channel sounder capable of capturing the \ac{CSI} with the corresponding user position and storing it into a database. The procedure is as follows: The mobile user transmits OFDM signals containing pilots and the embedded user position; the signal is received at the channel sounder using an antenna array, and the raw I/Q-samples are stored for further post-processing.
The complex channel coefficients per subcarrier and antenna are estimated offline and the position information is decoded from the payload, which is then saved into a database. 
To provide stable and consistent measurement datasets, the channel sounder needs to fulfill
the following requirements regarding range, signal quality and reproducibility. 
\begin{enumerate}
\item Requirement 1: Flexibility\\
Different frequency bands result in different propagation behavior and, thus, a basestation could be able to chose the best frequency band for each scenario \cite{3GPP2012}. This leads to the requirement that the channel sounder should be capable of operating in different frequency bands in the sub-\SI{6}{\giga \hertz} range. Moreover, different antenna geometries result in different channel estimation techniques as shown in \cite{6310934}. Therefore, the channel sounder should be independent of the antenna array used, which, also, should be freely configurable.
\item Requirement 2: Coverage\\
Coverage and, thus, keeping in mind the resulting path loss is key for each wireless communication system. LTE ``suburban'' was specified  to achieve a coverage of around \SI{5}{\kilo \metre}, with a maximum path loss of \SI{164.5}{\decibel} \cite{3GPP2012}. Since large parts of cellular systems are designed for a 1-2\ km coverage radius, the requirement for the channel sounder should be to cover a similar range. 
Taking into account the well-known breakpoint model \cite{3GPP2012}, a \SI{110}{\decibel} pathloss with an SNR of at least \SI{10}{\decibel} should still be achievable.
\item Requirement 3: Stability and reproducibility\\
The last requirement is that each estimated channel should be the most precise representation of the physical channel possible. 
Hardware impairments should not unnecessarily limit the measurement accuracy, e.g., a small \ac{ADC} resolution would lead to poor measurement accuracy. 
Moreover, the channel sounder should be stable over time and, thus, the channel must not change for a non-moving user, provided the propagation environment stays constant (e.g., indoor laboratory set-up).
\end{enumerate}
Considering these requirements for obtaining an accurate \ac{CSI} database, we designed a novel and low-cost channel sounder, as presented in the following.
\subsection{Principle of the channel sounder}

\begin{figure}
	\centering
	\includegraphics[width=0.5\textwidth]{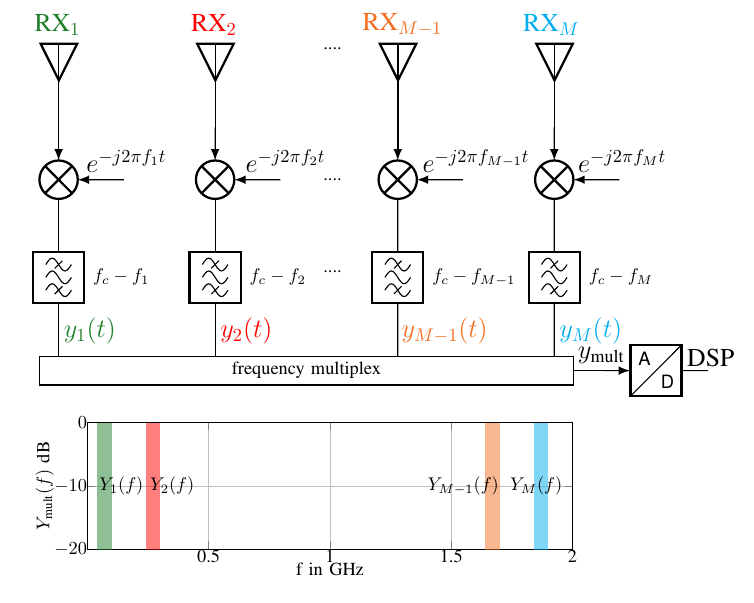}
	
	\caption{Basic principle of the channel sounder}
	\label{fig:Principle_Board}
	
\end{figure}

Fig. \ref{fig:Principle_Board} shows the basic principle of the novel channel sounder. Note that for simplification bandpass filter and amplifier for the \ac{AFE} are omitted. Each antenna signal is mixed from the carrier frequency $f_c$ onto an individual \ac{IF} $f_c - f_m$, $m=1,..,M,$ to become $y_m(t) \; \laplace \; Y_m(f)$ and is then combined using a low-loss frequency addition (as detailed in the next section). Here the number of antennas is defined as $M$. The combined signal $Y_{\rm mult}(f)$ is fed to a single \ac{ADC} with a high resolution and bandwidth.
After analog to digital conversion the \ac{DSP} unit splits the orthogonal frequency bands into single channels to estimate the CSI for each antenna separately (which can be done offline). 
As the discrete-time sampling is aligned in time among all antennas, a reduced number of calibration steps, e.g., no synchronization of multiple \ac{ADC}s or calibrating multiple transit times, is needed.
Although only the basic principle of receiver processing is shown, a transmitter (or even a full basestation) could be designed straightforwardly based on the same concept, simply by extending the \ac{AFE}. 
Note that, only the \ac{AFE} needs to be changed to work at different frequency bands; for example, our channel sounder is equipped for \SI{1.25}{\giga \hertz}, \SI{2.35}{\giga \hertz}  and \SI{3.75}{\giga \hertz} operation. 
%Later the remaining parts of Requirement 1 are fulfilled by using a selfbuild modular antenna array.

Concerning Requirement 2, and the effect of using only a single \ac{ADC}, the achievable \ac{SQNR} per antenna is computed. 
A number of $N_{\text{Ant}}=64$ antennas, each having a bandwidth of \SI{20}{\mega \hertz}, are combined. 
The signal is then clipped with a clipping range of $U_{\text{VPP}}=$\SI{100}{\milli \volt} and
quantized with different \ac{ENoB}.  

\begin{figure}
	\centering
	\includegraphics[width=0.5\textwidth]{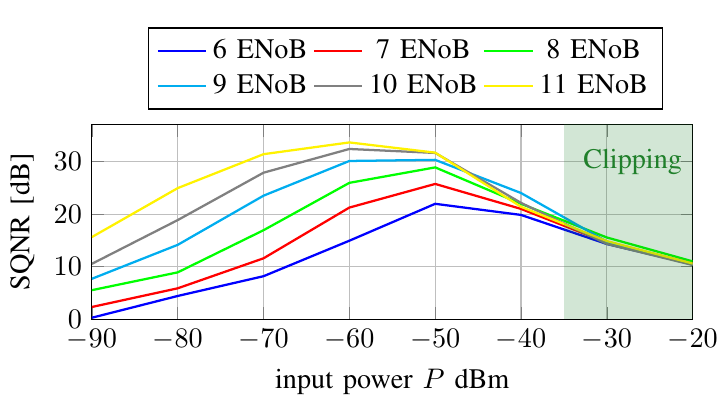}
	 \vspace*{-0.6cm}
	\caption{SQNR versus input Power at a 50Ohm input}
	\label{fig:SQNR_vs_Input_Power}
\end{figure}

Fig. \ref{fig:SQNR_vs_Input_Power} shows the simulated achievable \ac{SQNR} versus the input power. One can see that, the communication quality dependens on the \ac{ENoB}, as the subband signals are combined and each signal 
can only ``consume'' a fraction of the available \ac{ADC} resolution. 
For measuring an $x,y$-plane in space, a constant clipping range is preferable to maintain the channels' power relations.

\begin{figure}
	\centering
	\includegraphics[width=0.5\textwidth]{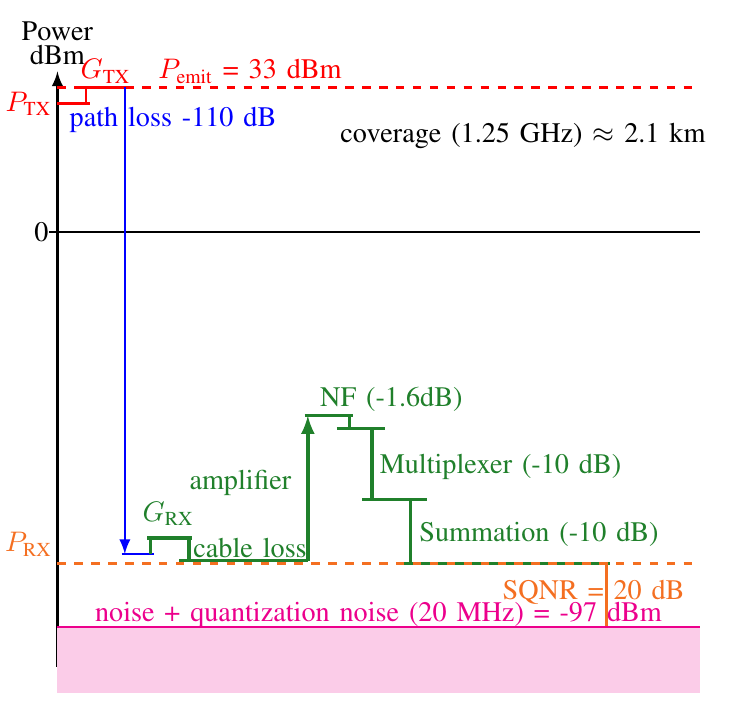}
	         \vspace*{-0.3cm}
	\caption{Link budget consideration}
	\label{fig:Linkbudget}
\end{figure}

Fig. \ref{fig:Linkbudget} shows the link budget consideration for this specific setup for one antenna chain. With an oscilloscope such as the LeCroy HD9404 \cite{HDO9404} with 7.8 \ac{ENoB}, and a targeted SNR of \SI{20}{\decibel}, the effective input power ranges from \SI{-77}{\decibel m} to \SI{-36}{\decibel m} for a bandwidth of \SI{20}{\mega \hertz}.
With our setup ($G_{\text{RX}}=\SI{3}{\decibel}$, \, $G_{\text{TX}}$=\SI{3}{\decibel}, \, $P_{\text{TX}}$=\SI{33}{\decibel},\, $\text{Amp}_{\text{RX}}$=\SI{20}{\decibel}) and the breakpoint model assumption ($h_{\text{TX}}=h_{\text{RX}}=\SI{1}{\metre}$) the resulting maximum coverage radius (addressing Requirement 2) computes to
\begin{equation}
d_{\text{max}}=\left(\frac{\sqrt{h_{\text{TX}}h_{\text{RX}}}}{10^{-\text{PL}_{\text{BP}}/40}}\right) \approx \SI{2.1}{\kilo \metre}.
\end{equation}
Moreover, we have decided to use an oscilloscope with 4 synchronized \ac{ADCs} and therefore the antennas are split onto 4 different channels, thus allowing to improve the SQNR per measurement chain.
Note that the effect of clipping could be mitigated by an \ac{AGC} unit, however, at the cost of losing the individual channel power relations that make up the measured multiple antenna channel vector. As expected, a coarse gain adjustment, equal for all RX antenna chains, is still needed and indeed performed prior to each scenario measurement.

\subsection{Realization of the channel sounder}

\begin{figure}
	\centering
	\includegraphics[width=0.5\textwidth]{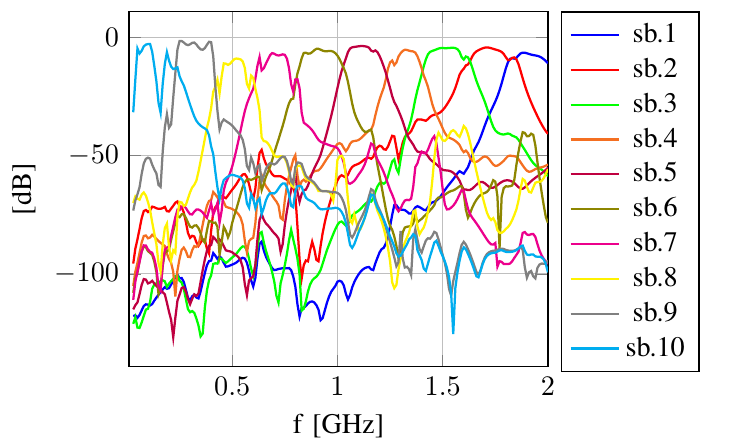}
	\vspace*{-0.5cm}
	\caption{Transfer functions of the multiplexing scheme}
	\label{fig:Transferfunction}
\end{figure}

As the straightforward addition of antenna signals with Wilkinson combiners \cite{1124668} would result in a loss of $3\log_2{\left(N_\text{Ant}\right)}$, a low-loss frequency multiplexing scheme is used. 
Due to size constraints and difficulties of matching many antennas on a \ac{PCB}, we have decided to place 10 antennas per \ac{PCB}.
For scaling to a higher number of antennas a production of the same \ac{PCB} with the same filters and adjusting only the \ac{IF} by programming the corresponding mixers is favorable. 
Therefore, a frequency-interleaved structure of combining \ac{PCB}s is used, resulting in 10 subbands (sb.) as shown in Fig. \ref{fig:Transferfunction}. 

Fig. \ref{fig:Transferfunction} indicates (simulation) that the subbands are separated by more than 20dB and, therefore, mutual cross-correlations should be sufficiently suppressed. Note that, in the example given, 4 antennas (4 \ac{PCB}s) are placed within one subband.  

\begin{figure}
	\centering
	\includegraphics[width=0.5\textwidth]{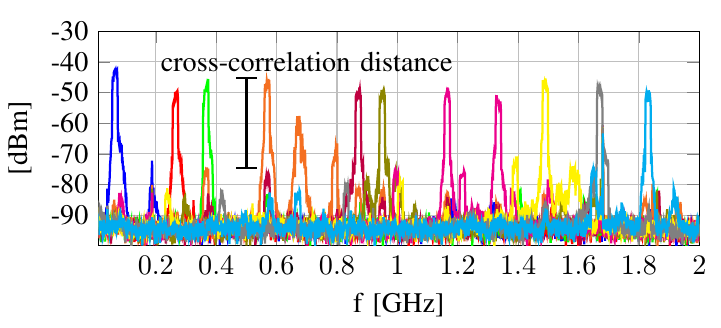}
	
	\caption{Measured cross-correlation power among subbands}
	\label{fig:Crosscorrelation}
\end{figure}

Fig. \ref{fig:Crosscorrelation} shows, for one specific (yet typical) example 
%(it was ensured that for every other frequency setting the chains are also uncorrelated) 
that the {\em measured} channels (and individual subbands) are decoupled with at least 30dB.
Therefore, any cross-correlation showing up using this architecture can be neglected in the following. 

\begin{figure}
\centering
\includegraphics[width=0.35\textwidth]{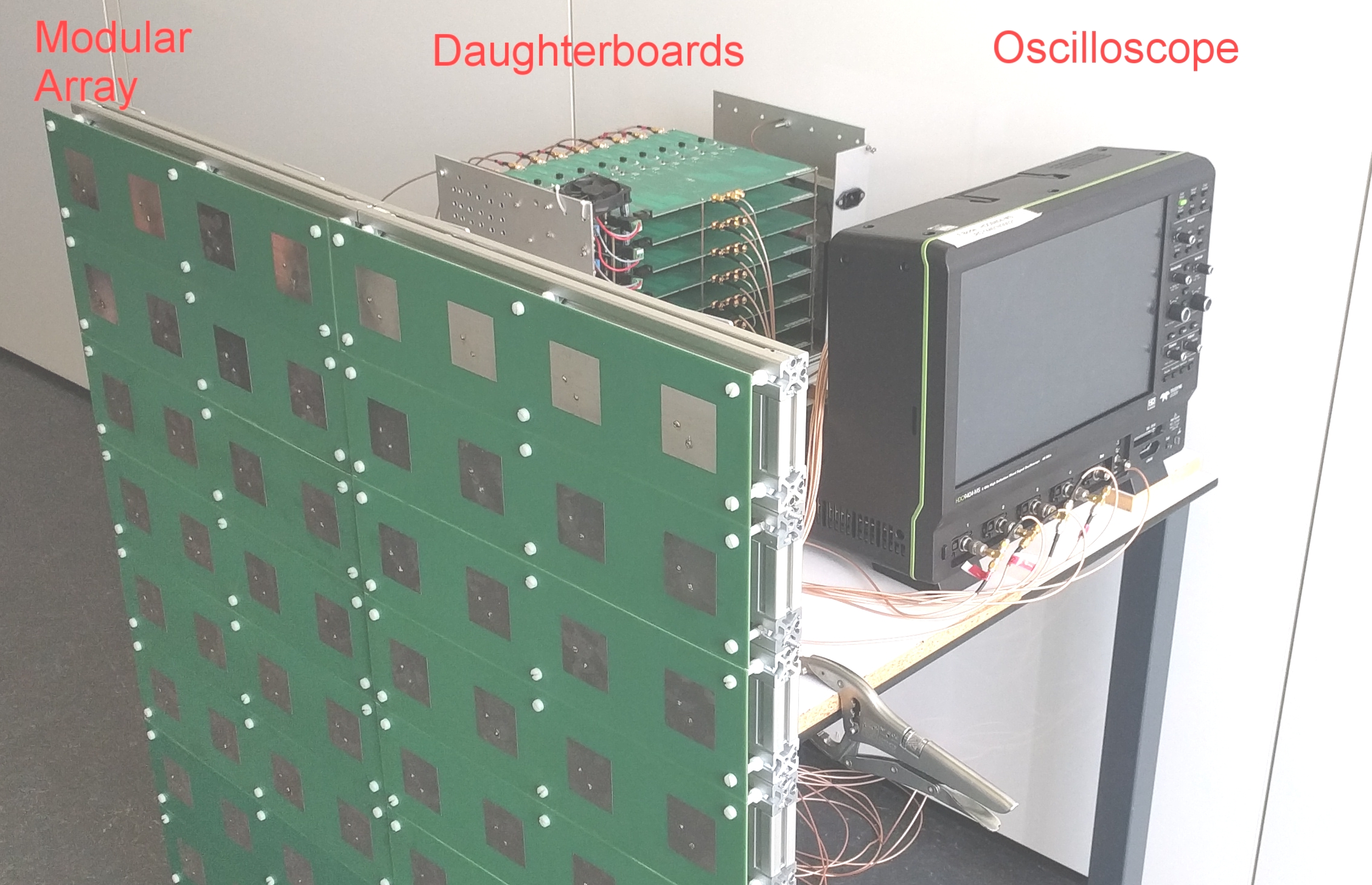}
%\subfloat[Blockdiagramm of 1 Antenna chain]{\input{tikz/Principle_of_Measurement.tikz}}\\
\caption{Channel sounder (daughterboards) with digital oscilloscope and antenna array}
\label{fig:MeasurementSetup}
\vspace*{-0.3cm}
\end{figure} 

Fig. \ref{fig:MeasurementSetup} shows the receiver part of the channel sounder with 64 antennas. The signal is received over the modular (dual polarized) array and then filtered, amplified, mixed (down-converted) and combined within so called ``daughterboards''. Then, two daughterboards with different frequency settings are combined onto one port of the oscilloscope. 
The signal is then mixed to baseband (offline, in digital signal processing), filtered, resampled and, finally, the channel coefficients are estimated. Note that the clock of an external source is fed to the oscilloscope and to all of the mixers on the daughterboards, resulting in a frequency aligned receiver.

The transmitter consists of a USRP \cite{USRP}, an amplification circuit and a dipole antenna, which transmits \ac{OFDM} symbols with a bandwidth of \SI{20}{\mega \hertz}, and 1024 subcarriers as pilots and data on the HAM radio frequency of \SI{1.25}{\giga \hertz}. Note that 10\% of the subcarriers are used as a guard band and the cyclic prefix was 1/8 of the OFDM symbol duration. For subcarrier modulation, a simple QPSK constellation was chosen.

\section{Verification of the Channel Sounder}
To satisfy Requirement 3, different stability and hardware impairment criteria are studied via test measurements. First, two fixed positions are measured to verify the stability of the channel sounder. Later, different indoor scenarios were chosen for generating a data set to evaluate user positioning algorithms based on deep learning.

\begin{figure*}
	\centering
	\includegraphics[width=1\textwidth]{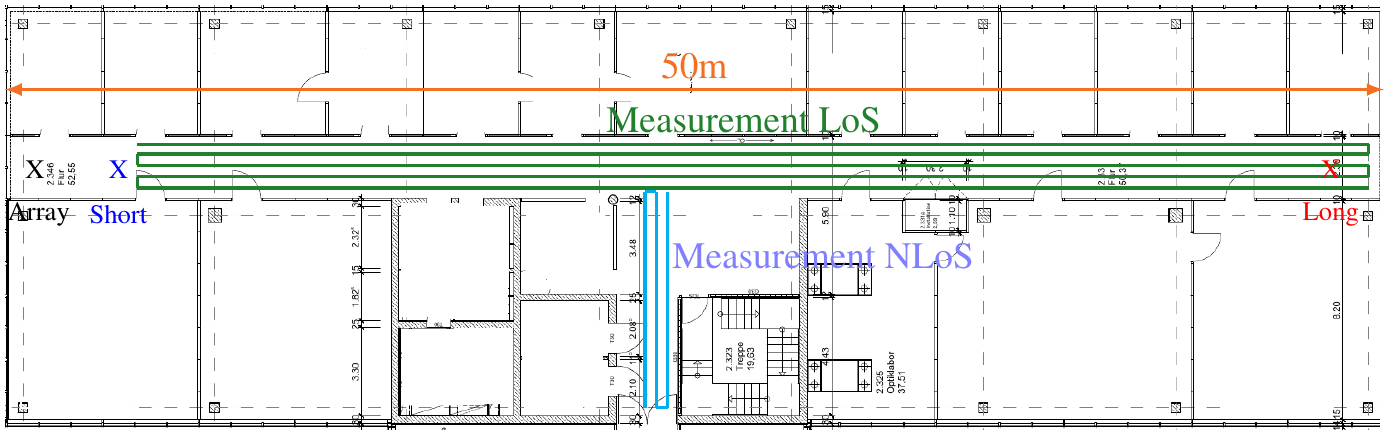}
	\caption{Indoor measurement scenario (office space with hallway)}
	\label{fig:Measurement_Scenario}
	\vspace*{-0.3cm}
\end{figure*} 

Fig. \ref{fig:Measurement_Scenario} shows the different indoor measurement scenarios where the array is configured as an 8x8 horizontally polarized rectangular array. 
First, two stable positions ``Short'' and ``Long'' are measured for stability study.
Later, the \ac{LoS} and \ac{NLoS} paths are measured at 3 different heights above the floor (0.5m, 1m, 1.5m), resulting in about \SI{600}{\metre} and \SI{200}{\metre} total measurement path length for \ac{LoS} and \ac{NLoS}, respectively.
As references for positioning, every \SI{2}{\metre} a grid point was marked on the floor, and positions were interpolated within.
It was verified that the resolution of this ``position ground truth'' is better than 10cm.

\subsection{Stability}
For the stability over time the correlation coefficient
\begin{equation}
\delta_{h}\left(\Delta t\right)=\frac{1}{N_{\text{Sub}}} \sum_{k=0}^{N_{\text{Sub}}}\frac{\left\Vert {\textbf{h}(k)}_{t}{\textbf{h}(k)}_{t+\Delta t}^{H}\right\Vert _{2}}{\left\Vert {\textbf{h}(k)}_{t}\right\Vert_{2}\left\Vert {\textbf{h}(k)}_{t+\Delta t}\right\Vert _{2}},\label{eq:ChannelCorrDef}
\end{equation}
of the $64 \times 1$ complex channel vector $\textbf{h}(k)$ is evaluated. 
The correlation coefficient is averaged over all subcarriers $k$. 
Moreover, the \ac{SNR} of each antenna is calculated over the \ac{EVM}.
\begin{figure}
\centering
\includegraphics[width=0.5\textwidth]{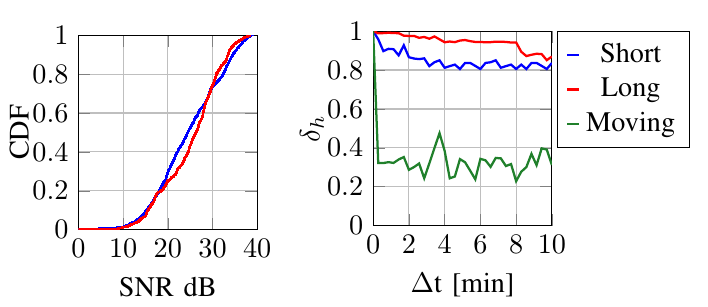}
%\subfloat[Correlation over Time]{\input{tikz/Correlation_over_time.tikz}}
\caption{Verifying amplitude/phase stability}
\label{fig:Stability_Investigation}
\vspace*{-0.3cm}
\end{figure}

Fig. \ref{fig:Stability_Investigation} shows the correlation coefficient over time for the different measurements ``Short'' , ``Long'' and ``Moving'', and plots the \ac{CDF} of all measured \ac{SNR}s. 
It can be seen that the SNR always stays above \SI{10}{\decibel} for every antenna and, on average is about \SI{25}{\decibel}. Therefore, the signal quality over the 10\ minute measurement is good enough to verify  stability. 
With an average correlation coefficient above 0.8 over 10\ min, the measurement setup can be considered to be sufficiently stable.  To cross-check, when moving the transmitter, the correlation coefficient falls well below 0.5, which is a plausible observation.

\subsection{Sanity checks for hardware impairments}
For a transmission scheme with one transmitter and 64 receive antennas, the received signal can be written \cite{Bjrnson2017} as 
\begin{equation}
\bf{y}_{\it{k}} = \bf{w}_{\it{k}}^H\bf{h}_{\it{k}}s_{\it{k}} + \bf{w}_{\it{k}} \bf{n}_{\it{k}},
\end{equation}
where $\bf{w}_{\it{k}}$ is the $1\times 64$ linear combining vector, $\bf{h}_{\it{k}}$ is the $64\times 1$ channel vector, $s_{\it{k}}$ is the transmitted QPSK symbol and $\bf{n}_{\it{k}}$ is the $64\times 1$ additive white Gaussian noise vector for each subcarrier $k$.
To verify whether the hardware impairments are limiting the gain of the linear combining scheme, a simple \ac{MRC} scheme is investigated, where the combining vector
\begin{equation}
 \bf{w}_{\it{k}} = \bf{\hat{h}}_{\it{k}},
\end{equation}
is used and where $\bf{\hat{h}_{\it{k}}}$ is the estimated channel vector. In a non-hardware limited system and with equal \ac{SNR} per antenna, a 3dB gain in the \ac{BER} curve per doubling of the number of antennas would be expected. 

\begin{figure}[]
	\centering
	\includegraphics[width=0.5\textwidth]{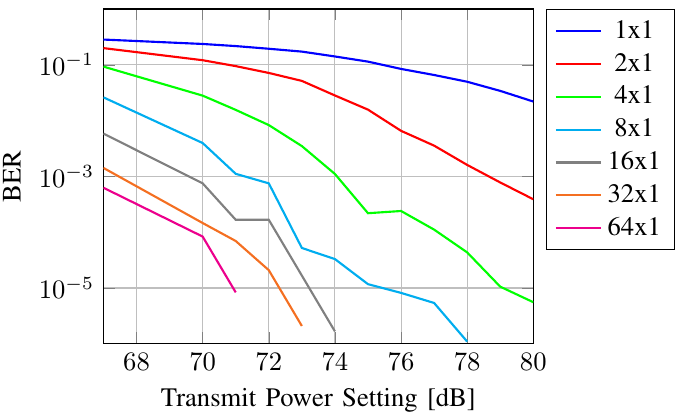}
	
	\caption{Measured \ac{MRC} gain vs number of RX antennas}
	\label{fig:MRC_BER}
	\vspace*{-0.3cm}
\end{figure}
Fig. \ref{fig:MRC_BER} shows the \ac{BER} curves for different number of antennas, when the transmitter gain of the TX-USRP is sweeped. Hereby QPSK symbols were transmitted and the \ac{BER} of a channel were measured after combining.
Note that all 64 antennas were measured and the ``virtual'' antennas to combine were chosen as follows: An antenna with a moderate to low SNR was chosen as the first BER curve and then the antenna with the next higher SNR was used for \ac{MRC}, and so on. 
With the difference in SNR a gain higher than \SI{3}{\decibel} can occur, as can be seen in the figure (channel is not ideal AWGN). 
Moreover, the expected gains of \ac{MRC} appear and verify that the hardware impairments  constrain our system only very little (even though the gain from 32 antennas to 64 antennas is not quite 3 dB, indicating the presence hardware impairments), verifying the usability of the channel sounder. 
% or channel...

\section{First Trials for Indoor 3D-User Positioning}
One first example for showing the potential of position-tagged CSI dataset is shown next, for the case of the yet unsolved problem of indoor user positioning \cite{Lymberopoulos2018}. The two paths (\ac{LoS} and \ac{NLoS}) shown in Fig. \ref{fig:Measurement_Scenario} are 
measured at 3 different heights (\SI{0.5}{\metre}, \SI{1}{\metre}, \SI{1.5}{\metre}) above the floor. 
Each time a position-tagged dataset of about 2000 spatial points per height is created, resulting in an average distance per point of $\lambda/2$.

\begin{table}			
\centering
\caption{Layout of the neural net for 64 antennas}
\begin{tabular}{c|c}
Layers:                           & Output dimensions           \\
\hline                                                                       
Input                                  & 64 x 922 x 2 (Re/Im)   \\
Conv2D                       & 64 x 922 x 32                  \\
AveragePooling                  & 64 x 231 x 32                  \\
Conv2D                      & 64 x 231 x 32                  \\
AveragePooling                  & 64 x 58  x 32                   \\
Dense                     & 256                           \\
Dense                       & 256                           \\
Dense                      & 256                        \\
Dense                           & 3 (X/Y/Z)                 \\
\end{tabular}
\label{tab:nn_layout2}
%\vspace*{-0.3cm}
\end{table}

An \ac{NN} (example given for 64 antennas in Table \ref{tab:nn_layout2}) is trained, based on the measured channels from all antennas and on all sub-carriers, to estimate the position. For this, 90 \% percent of the data was used for training and 10 \% for validation. Note that the \ac{NN} has a number of weights ranging from $\approx1,000,000$ to $\approx16,000,000$.
This is due to the fact that for a fair comparison the net structure is the same when going from 2 to 64 antennas,  but the input dimension increases. The convolutional layers were chosen according to the rationale that always 4 subcarriers are assumed to be correlated.
This information is averaged for condensing the data without losing information.

\subsection{Line-of-Sight Positioning}

\begin{figure}
	\centering
	\includegraphics[width=0.5\textwidth]{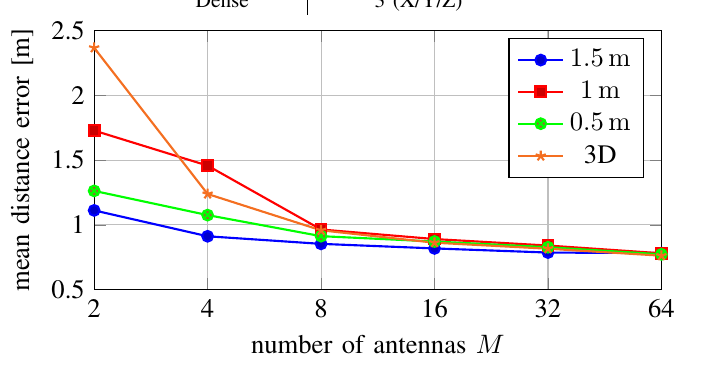}
	\vspace*{-0.6cm}
	\caption{Measured position error, indoor \ac{LoS} scenario}
	\label{fig:IndoorPos}
	\vspace*{-0.3cm}
\end{figure}
Fig. \ref{fig:IndoorPos} shows the mean distance error between the actual position $\mathbf{d}$ and the expected position $\hat{\mathbf{d}} $ over the number of antennas.
As can be seen, when going from 2D to 3D user localization a difference only occurs for the region of small number of antennas. 
A positioning precision of about \SI{75}{\centi \metre} is achieved over the entire measurement area, outperforming other state-of-the-art methods \cite{Lymberopoulos2018}. 
To better understand the further possible improvements, the histogram of the positioning error is further examined.

\begin{figure}
	\centering
	\includegraphics[width=0.5\textwidth]{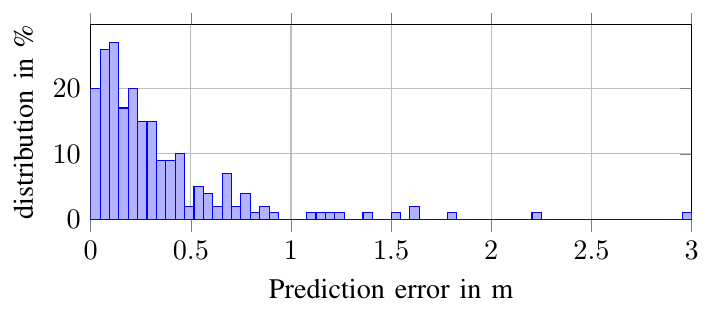}
	\vspace*{-0.6cm}
	\caption{Distribution of distance error}
	\label{fig:Histogram_of_Error}
	\vspace*{-0.3cm}
\end{figure}

Fig. \ref{fig:Histogram_of_Error} shows the histogram of the prediction error which follows a Rayleigh-like distribution, as the $x,y,z$ error seems to be Gaussian distributed. Moreover, there are positions having an error of meters, which were located at sharp edges nearby doors.
For the \ac{LoS} case, passive radar systems would also be a possible solution, but in the more sophisticated case of \ac{NLoS} they do not work. This is the reason why we move on to studying the \ac{NLoS} region more closely.

\subsection{Non-Line-of-Sight Positioning}

\begin{figure}
	\centering
	\includegraphics[width=0.5\textwidth]{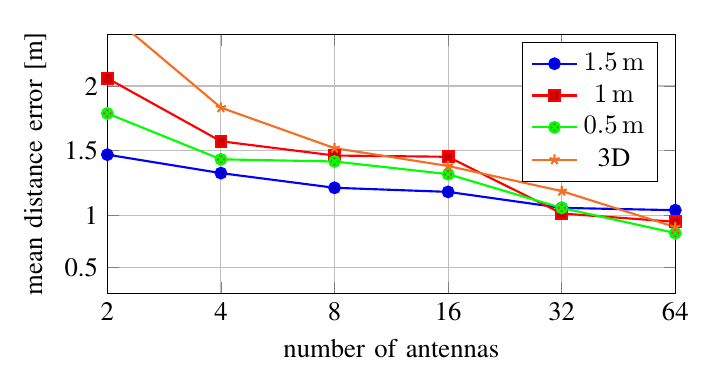}
	\vspace*{-0.6cm}
	\caption{Measured positioning error, indoor NLoS scenario}
	\label{fig:IndoorPos_NLoS}
	\vspace*{-0.3cm}
\end{figure}

Fig. \ref{fig:IndoorPos_NLoS} shows the positioning error over the number of receive antennas. The prediction error increases to ~\SI{95}{\centi \metre}, which is due to both the \ac{NLoS} environment as well as the further distance from the antenna array. Still, the system predicts the user positions remarkably well.
%Therefore indoor user position is enabled by Massive MIMO and the corresponding datasets.
It appears that \ac{DL} in combination with Massive MIMO datasets has the potential of addressing the task of indoor user localization with better prediction accuracy than other known methods thus far.

\section{Outlook and Conclusion}
We introduce a channel sounder that enables the generation of a publicly available position-tagged Massive MIMO database, in support of better evaluating channel properties as well as studying deep learning techniques on measured channel data.
The main requirements regarding stability, flexibility and coverage for a channel sounder where shown.
A novel channel sounder architecture is proposed to meet those requirements, and its viability is verified via simulation as well as actual measurements. 
%Measurements show the stability of the build channel sounder and that hardware impairments are not limiting the system performance.
%
Using this channel sounding data, initial results for 3D-indoor user positioning via deep learning have been obtained, beating state of the art systems, even in the more difficult scenario of \ac{NLoS}. 
Further work will address different frequency bands, other antenna geometries and
propagation scenarios. In the near future and enriched with new measurements a database will be made public for the scientific community to
test-run their own algorithms for various applications, beyond precoding and user positioning.

\bibliographystyle{IEEEtran}
\bibliography{MassiveMIMO_Basestation.bib}
\end{document}